\documentclass[aps,prb,amsmath,reprint,superscriptaddress,longbibliography]{revtex4-1}
\usepackage{amsmath,txfonts,bm,dcolumn,graphicx,color}
\usepackage{physics}
\begin{document}

\title{Visualizing Complex-Valued Molecular Orbitals}
\author{Rachael Al-Saadon}
\affiliation{Department of Chemistry, Northwestern University, 2145 Sheridan Rd., Evanston, IL 60208, USA.}
\author{Toru Shiozaki}
\email{shiozaki.toru@gmail.com}
\affiliation{Department of Chemistry, Northwestern University, 2145 Sheridan Rd., Evanston, IL 60208, USA.}
\author{Gerald Knizia}
\email{knizia@psu.edu}
\affiliation{Department of Chemistry, The Pennsylvania State University, 401A Chemistry Building, University Park, PA 16802, USA.}
\date{\today}

\begin{abstract}
We report an implementation of a program for visualizing complex-valued molecular orbitals. The orbital phase information is encoded on each of the vertices of 
triangle meshes using the standard color wheel.
Using this program, we visualized the molecular orbitals for systems with spin--orbit couplings, external magnetic fields, and complex absorbing potentials.
Our work has not only created visually attractive pictures, 
but also clearly demonstrated that the phases of the complex-valued molecular orbitals carry rich chemical and physical information of the system,
which has often been unnoticed or overlooked.
\end{abstract}

\maketitle

\section{Introduction}
The phase of molecular orbitals is fundamentally important in predicting and understanding chemical reactions.
One of the earliest examples in the literature is the demonstration that
the selectivity of the Diels--Alder reactions can be explained by the phases of the highest-occupied and lowest-unoccupied
molecular orbitals.\cite{Woodward1965JACS} 
Molecular orbitals have been ever since considered as a key descriptor for chemical reaction mechanisms. 
Many of the research articles in computational chemistry thus report the graphic pictures of molecular orbitals, which are often
obtained by the standard visualization softwares such as Molden,\cite{molden} Avogadro,\cite{avogadro} IQMol,\cite{iqmol} IboView\cite{Knizia2013JCTC,Knizia2015ACIE} to name a few.

To the best of our knowledge, visualization of molecular orbitals (and associated properties) has been thus far limited to
that of real-valued orbitals due to the lack of implementation.
This is partly because the software for real-valued orbitals does suffice for traditional electronic structure simulations in which the non-relativistic time-independent
Schr\"{o}dinger equation is solved for bound states with open boundary conditions, owing to the fact that the Hamiltonian for such systems is real
and the phase of the orbitals can be chosen to be either $+1$ or $-1$.
There are, however, various situations in which the molecular orbitals become complex, including simulation of systems
with relativistic\cite{Reiherbook}/magnetic\cite{Tellgren2008JCP,Reynolds2015PCCP}
contributions, that with periodic\cite{Manbybook}/absorbing\cite{Riss1993JPB,Jagau2017ARPC} boundary conditions, and that with explicit time dependence.\cite{Goings2017WIREs} 
Another example is complex generalized Hartree--Fock wave functions that can describe non-coplaner spin polarization.\cite{Fukutome1981IJQC, Henderson2018JCTC}
Since computational tools for such systems are becoming widely available in recent years, it is of great importance to be able to visualize such orbitals. 

In this work, we have developed a new computer program that allows for visualizing complex-valued molecular orbitals (and other properties). 
We apply this program to systems under an external magnetic field, systems with spin--orbit coupling, and systems with 
the absorbing boundary condition.
Numerical examples will be shown to demonstrate surprisingly rich information that is pertained in the phase of those orbitals.

\section{Technical details}
\begin{figure}[b]
\includegraphics[keepaspectratio,width=0.20\textwidth]{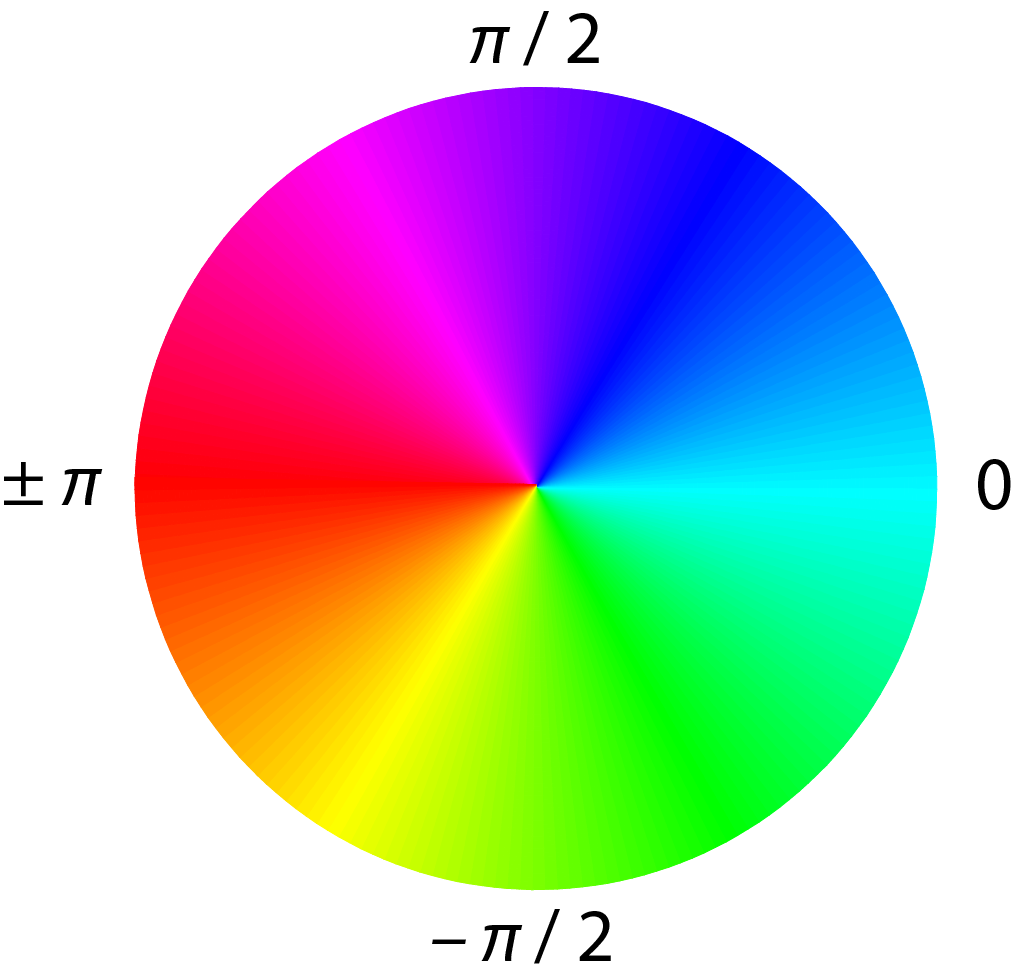}
\caption{Color wheel that represents the phases.\label{colorw}}
\end{figure}
To compute the isosurface of a complex-valued molecular orbital, we first take the norm of the orbitals and reuse the components in the IboView programs\cite{Knizia2013JCTC,Knizia2015ACIE} that have been written for real-valued molecular orbitals.
To do this, we simply evaluate the complex orbitals in real space and take the norm at each point:
\begin{align}
\label{comp1}
|\psi_i(\mathbf{r})| = \left|\sum_\mu C_{\mu i} \phi_\mu(\mathbf{r})\right| 
\end{align}
where $\psi_i$ is an $i$-th molecular orbital, $\phi_\mu$ are basis functions that are real valued, and $C_{\mu i}$ are the molecular-orbital coefficients that are complex.
We use the so-called Marching Cubes algorithm\cite{Lorensen1987ACM} for computing the triangular (polygon) meshes of the orbital isosurface.
Following the default setting with IboView, the isosurfaces are constructed such that 80\% of the molecular orbitals are encapsulated within the surface.

The normals of the isosurfaces are computed exactly from the derivative of the wave functions at each of the vertices.
This has also been implemented in IboView and reused in this work.\cite{Knizia2013JCTC,Knizia2015ACIE}
The derivative of the norm can be computed as 
\begin{subequations}
\label{comp2}
\begin{align}
&\partial_w |\psi_i(\mathbf{r})| =
\Re[\partial_w \psi_i(\mathbf{r})]\cos\theta(\mathbf{r}) + \Im[\partial_w \psi_i(\mathbf{r})] \sin\theta(\mathbf{r}),\\ 
&\partial_w \psi_i(\mathbf{r}) = \sum_\mu C_{\mu i} \partial_w \phi_\mu(\mathbf{r}),\\
&\theta(\mathbf{r}) = \mathrm{arg}[\psi_i(\mathbf{r})],
\end{align}
\end{subequations}
where $\partial_w = \partial/\partial w$ and $w = x$, $y$, and $z$.

Subsequently, we calculate the phase angle $\theta$ of the molecular orbital at each of the vertices of the triangular meshes.
The phase angle is then converted to a color code based on the RGB color wheel that maps the hue of the colors to [0, $2\pi$], as shown in Fig.~\ref{colorw}, 
or, more quantitatively,
\begin{align}
\label{cweq}
\begin{array}{lllll}
-3 \le p < -2 && R = 1 & G = p+ 3 & B = 0 \\
-2 \le p < -1 && R = -1-p & G = 1 & B = 0 \\
-1 \le p < 0 && R = 0 & G = 1 & B = p+1 \\
0 \le p < 1 && R = 0 & G = 1-p & B = 1 \\
1 \le p < 2 && R = p-1 & G = 0 & B = 1 \\
2 \le p < 3 && R = 1 & G = 0 & B = 3-p \\
\end{array}
\end{align}
with $p = 3\theta /\pi$. 
Here the RGB scale takes the values between 0 and 1.
The resulting color codes are added to the vertex information before the surfaces are rendered.
Note that the color wheel is routinely used for visualizing complex functions in standard software programs, e.g.,  Mathematica.\cite{Mathematica}

The above algorithms are implemented using the locally-modified IboView program. The computation of Eqs.~\eqref{comp1} and \eqref{comp2} are performed in the modified version of the
BAGEL programs.\cite{bagel,Shiozaki2018WIREs}

\section{Examples}
\subsection{Molecules in external magnetic fields}
\begin{figure}[t]
\includegraphics[keepaspectratio,width=0.48\textwidth]{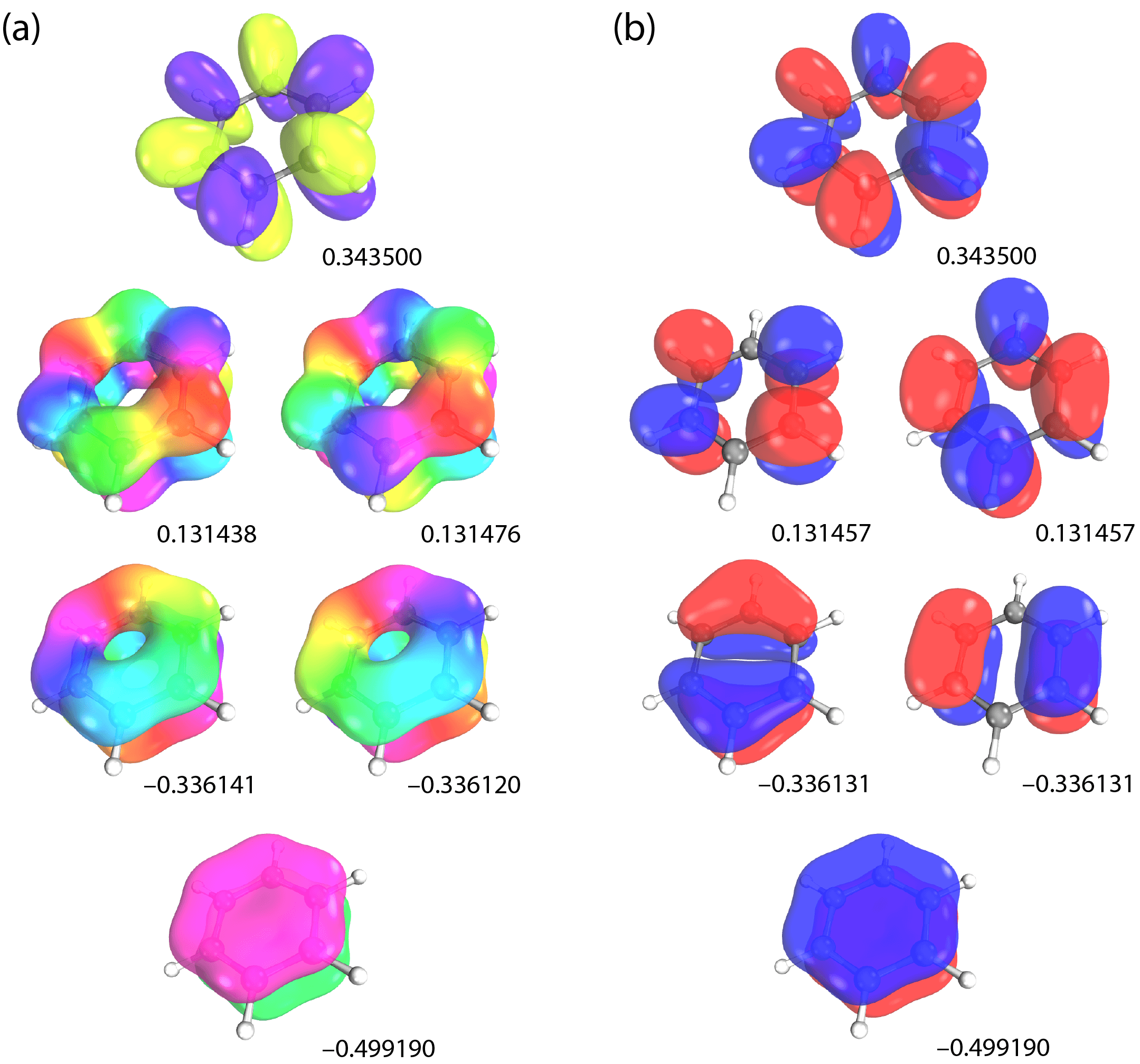}
\caption{$\pi$ orbitals of benzene placed in the $xy$-plane (a) with and (b) without external magnetic fields (5~T in the $z$ direction).
For coloring in (a), see Fig.~\ref{colorw}. (b) is depicted using the code for real-valued orbitals.
The numbers are the orbital energies in Hartrees.
\label{benzenemag}}
\end{figure}

When molecules are placed under an external magnetic field, the wave function becomes 
complex, because the momentum operator in the Hamiltonian is modified as
\begin{align}
\hat{\mathbf{p}} \to \hat{\boldsymbol \pi} = -i \nabla + \mathbf{A}
\end{align} 
where $\mathbf{A}$ is a vector potential generated by the external magnetic field.
Furthermore, one often uses the so-called gauge-including atomic orbitals (GIAO) to 
retain gauge-origin invariance,\cite{London1937JPR,Wolinski1990JACS}
\begin{align}
\phi_i(\mathbf{r}) \to \phi'_i(\mathbf{r}) = \exp\left[-i\mathbf{A}(\mathbf{R})\cdot \mathbf{r} \right] \phi_i(\mathbf{r})
\end{align}
where $\mathbf{R}$ is the location of the atom that the basis function $i$ belongs to.
Note that the use of GIAO-based programs for studying systems under an external magnetic field is relatively new, pioneered by Tellgren and co-workers,\cite{Tellgren2008JCP}
who have reported interesting chemical applications in astrochemistry,\cite{Lange2012S,Stopkowicz2018IJQC} 
followed by one of the authors.\cite{Reynolds2015PCCP,Reynolds2018JCP} 
In this section, we used the GIAO-based Hartree--Fock program in the BAGEL package\cite{bagel} as described in Ref.~\onlinecite{Reynolds2015PCCP}. 

In Fig.~\ref{benzenemag}, we compare the $\pi$ orbitals of a benzene molecule with and without an external magnetic field.
The benzene molecule is at its equilibrium geometry on the $xy$-plane, and a magnetic field of 5~T was applied in the $z$ direction. 
The def2-SVP basis set and corresponding fitting basis sets were used.\cite{Schafer1992JCP}
In the absence of magnetic fields, there are two sets of doubly-degenerate $\pi$ orbitals (one bonding and one anti-bonding orbitals). 
The degenerate sets become split when the magnetic field is applied; this is because the magnetic field favors counter-clockwise electronic current
in the direction of the field. Therefore, the degenerate occupied orbitals take linear combinations such that the phase changes $\pm 2n\pi$ when rotating around the principal axis (where
$n$ is the number of nodes perpendicular to the $xy$-plane for the orbitals in the absence of magnetic fields).
The hybridization phenomena is clearly depicted in the figures, facilitating intuitive understanding of the molecular orbitals in the 
presence of the external magnetic field.
The amount of phase changes $\pm 2n\pi$ and $\pm 4n\pi$ cannot be read from the phaseless counterparts.

\subsection{Spin--orbit coupled wave functions}
\begin{figure}[tb]
\includegraphics[keepaspectratio,width=0.45\textwidth]{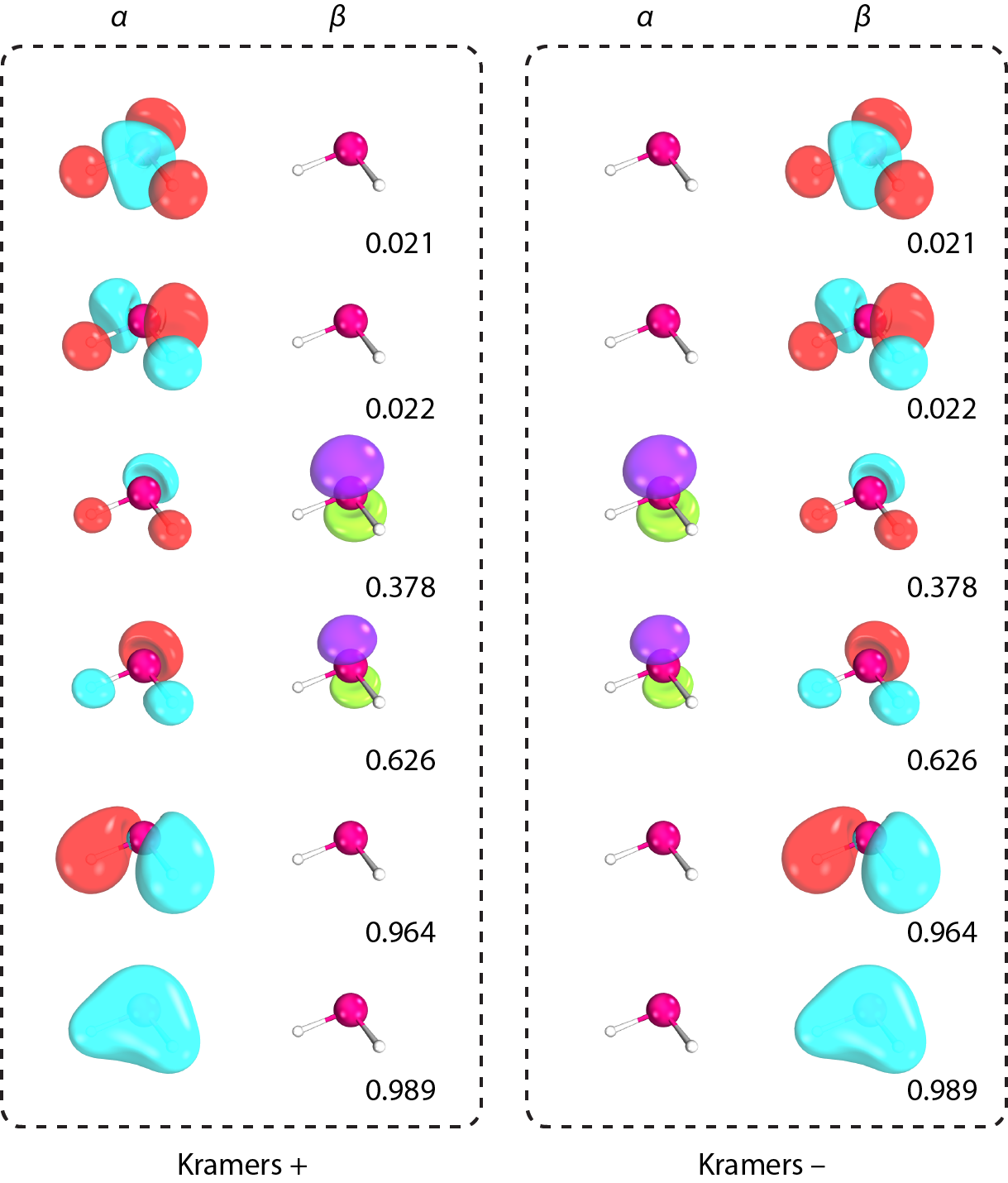}
\caption{The natural orbitals of one of the spin--orbit coupled states for PbH$_2$ obtained by the state interaction method.
The numbers denote orbital occupation.
The hue of the colors represents the phase of the orbitals (see Fig.~\ref{colorw}).
\label{pbh2fig}}
\end{figure}

One of the approaches to taking into account spin--orbit couplings is to use the so-called state interaction method.\cite{Malmqvist2002CPL,Sayfutyarova2016JCP}
In this approach, one first obtains a number of pure spin states in the absence of spin--orbit couplings, which are then used as a basis to construct an effective Hamiltonian that includes
spin--orbit couplings. The spin--orbit coupling elements are complex: for instance, the one-body part of the spin--orbit coupling matrix elements from the Breit--Pauli Hamiltonian are\cite{Reiherbook}
\begin{align}
(H^\lambda_\mathrm{SO,1})_{pq} = -\frac{i}{4c^2} \sum_{A} \epsilon_{\mu\nu}^\lambda
\mel{\frac{ \partial \phi_p}{\partial \mu}}{\frac{Z_A}{r_A}}{\frac{\partial \phi_q}{\partial\nu}}
\end{align}
where the left-hand side are the matrix elements associated with the Pauli operator $\hat{\sigma}^\lambda_{pq}$, $\lambda$, $\mu$, and $\nu$ label Cartesian components, $c$ is the speed of light, $\epsilon_{\mu\nu}^\lambda$ is the Levi--Civita symbol, $p$ and $q$ label basis functions, and $A$ labels nuclei. 

As a simple example, we considered PbH$_2$ that exhibits spin--orbit coupling of about $2500$~cm$^{-1}$,\cite{Matsunaga1996JCP} which is a heavy-element analogue of the carbene biradical. 
The geometry was set to $r=1.880$~{\AA} and $\theta=91.5^\circ$ with the $C_{2v}$ symmetry.
One singlet state and one triplet state were included in the effective Hamiltonian.
We first performed a state-averaged CASSCF calculation with the spin-free DKH2 Hamiltonian,\cite{Hess1986PRA} averaging over both the singlet and triplet states. 
The full-valence active space (6 electrons in 6 orbitals) and the ANO-RCC basis set\cite{Roos2004JPCA} were used.
Subsequently, the spin-nonconserving density matrices associated with the Pauli operator $\hat{\sigma}^\lambda_{pq}$ were computed.
The effective Hamiltonian was then formed and diagonalized to obtain the spin--orbit coupled states (total of 4 spin--orbit states).
In Fig.~\ref{pbh2fig}, we present the natural orbitals for one of the triplet states that is coupled to the singlet state (i.e., the state that is predominantly $|1,1\rangle + |1,-1\rangle$ using
the $|L,L_z\rangle$ notation).
From this figure, one can visually determine the orbitals that play an important role in spin--orbit interaction and their phase relationshiops. The third and fourth orbitals 
in each column are the hybridized singly occupied orbitals that mix $\alpha$ and $\beta$ components owing to the spin--orbit coupling.

\begin{figure}[tb]
\includegraphics[keepaspectratio,width=0.45\textwidth]{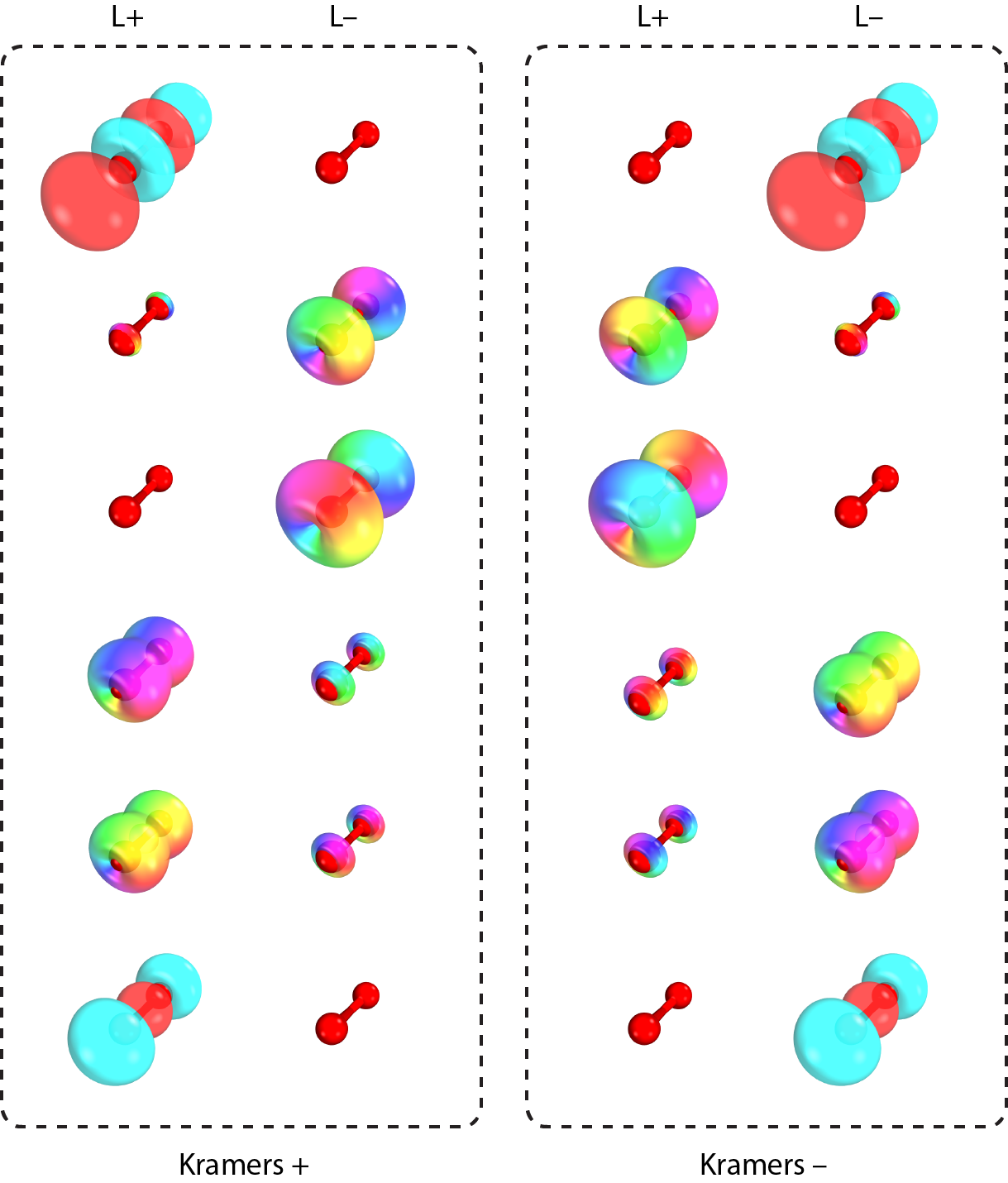}
\caption{The molecular spinors of O$_2$ consisting of $2p$ orbitals obtained by Dirac CASSCF.
The $L+$ and $L-$ components of the Kramers pairs are presented.
From top to bottom, the orbitals are the $\sigma^\ast$, $\pi^\ast \times 2$, $\pi \times 2$, and $\sigma$ orbitals.
The hue of the colors represents the phase of the orbitals (see Fig.~\ref{colorw}).
\label{o2mo}}
\end{figure}

Another approach to accouting for spin--orbit interaction is the four-component relativistic approaches that seek to directly solve the Dirac equations that include spin--orbit interaction a priori.\cite{Reiherbook}
The Dirac equation reads
\begin{subequations}
\begin{align}
& \hat{H} = \sum_{i} \hat{h}(i) + \sum_{i<j} \hat{g}(i,j),\\
& \hat{h}(i) = c^2 (\beta - I_4) + c(\boldsymbol{\alpha} \cdot \hat{\mathbf{p}}_i) - \sum_A^{\mathrm{atoms}} \frac{Z_A}{r_{iA}}, \label{1e Hamiltonian}
\end{align}
\end{subequations}
in which $\boldsymbol{\alpha}$ and $\beta$ are Dirac's 4$\times$4 matrices;
$I_4$ is a 4$\times$4 identity matrix;
$c$ denotes the speed of light; $\hat{\mathbf{p}}=-i\nabla$ is the momentum operator;
and, $\hat{g}(i,j)$ is the electron--electron interaction (in this work, we used the standard Coulomb operator) with $i$ and $j$ labeling electrons.
Such four-component approaches have become applicable to large molecules of chemical interest in the past decade.\cite{Kelley2013JCP,Storchi2013JCTC,Repisky2014JCTC}
One of the consequences of using the four-component wave function approaches is that the orbitals are to be represented by complex-valued 4-spinors;
therefore, our visualization code for complex-valued molecular orbitals are particularly useful for the analyses of four-component wave functions. 

As a simple example, the molecular spinors are presented for the triplet O$_2$ molecule computed by the Dirac CASSCF method
with density fitting as described in Ref.~\onlinecite{Reynolds2018JCP} (note that other implementations of Dirac CASSCF includes those reported in Ref.~\onlinecite{Jensen1996JCP,Thyssen2008JCP}).
The def2-SVP basis set and corresponding fitting basis sets were used.\cite{Schafer1992JCP}
CAS(2$e$,2$o$) was used and three states were averaged.
Figure~\ref{o2mo} shows the valence canonical orbitals of this system, in which
only the large components (L$+$ and L$-$) of the 4-spinors are included. 
Both Kramers $+$/$-$ orbitals are  shown.
Note that the corresponding Kramers $+$/$-$ orbitals are degenerate due to the time-reversal symmetry; therefore, rotations between them
are arbitrary.
It is shown that both bonding and anti-bonding $\pi$ orbitals are hybridized in the presence of spin--orbit coupling that breaks the degeneracy.
In addition, one can clearly observe the time-reversal relation between the Kramers $+$/$-$ pairs,
\begin{subequations}
\begin{align}
&\psi^+_{L+}(\mathbf{r}) = [(\psi^{-}_{L-}(\mathbf{r})]^\ast,\\
&\psi^+_{L-}(\mathbf{r}) = -[\psi^{-}_{L+}(\mathbf{r})]^\ast.
\end{align}
\end{subequations}
Visually confirming this time-reversal symmetry is only possible using visualization code for complex-valued orbitals with phases.
These relationships are especially non-trivial for hybridized orbitals;
when checking these relationships, notice in Fig.~\ref{colorw} that the phase angles 0 and $\pi$ are cyan and red, respectively.

\subsection{Molecules with metastable states}
The renewed interest in metastable states in the past decade has inspired the development of various advanced computational tools.\cite{Riss1993JPB,Jagau2017ARPC} 
When modeling metastable states, in which the bound-state wave functions are mixed with continuum wave functions, 
one often uses an effective Hamiltonian that truncates the continuum degrees of freedom in one way or another;
and this truncation leads to Hamiltonians whose eigenvalues are complex. The imaginary part of the eigenvalues is related to the lifetime of the state.
Since the associated wave functions are complex, analysis of excited-state wave functions
warrants visualization of complex-valued orbitals.
Usually, the real and imaginary parts of the orbitals are presented separately.\cite{Jagau2017ARPC}

Here we show an example that employs the so-called complex absorbing potential (CAP) using the second-order multireference perturbation theory,
which has been reported by Bravaya and co-workers\cite{Kunitsa2017JCP} using an uncontracted multireference perturbation theory called the XMCQDPT2 method\cite{Granovsky2011JCP}
(in this work, we use instead the fully contracted variant, XMS-CASPT2, as implemented in BAGEL\cite{MacLeod2015JCP,Vlaisavljevich2016JCTC}).
In short, this approach includes both dynamical electron correlation and complex absorbing potentials as perturbations in the post-CASSCF process. 
For simplicity, the natural orbitals are computed from the CASSCF part of the wave functions.

\begin{figure}[tb]
\includegraphics[width=0.45\textwidth,keepaspectratio]{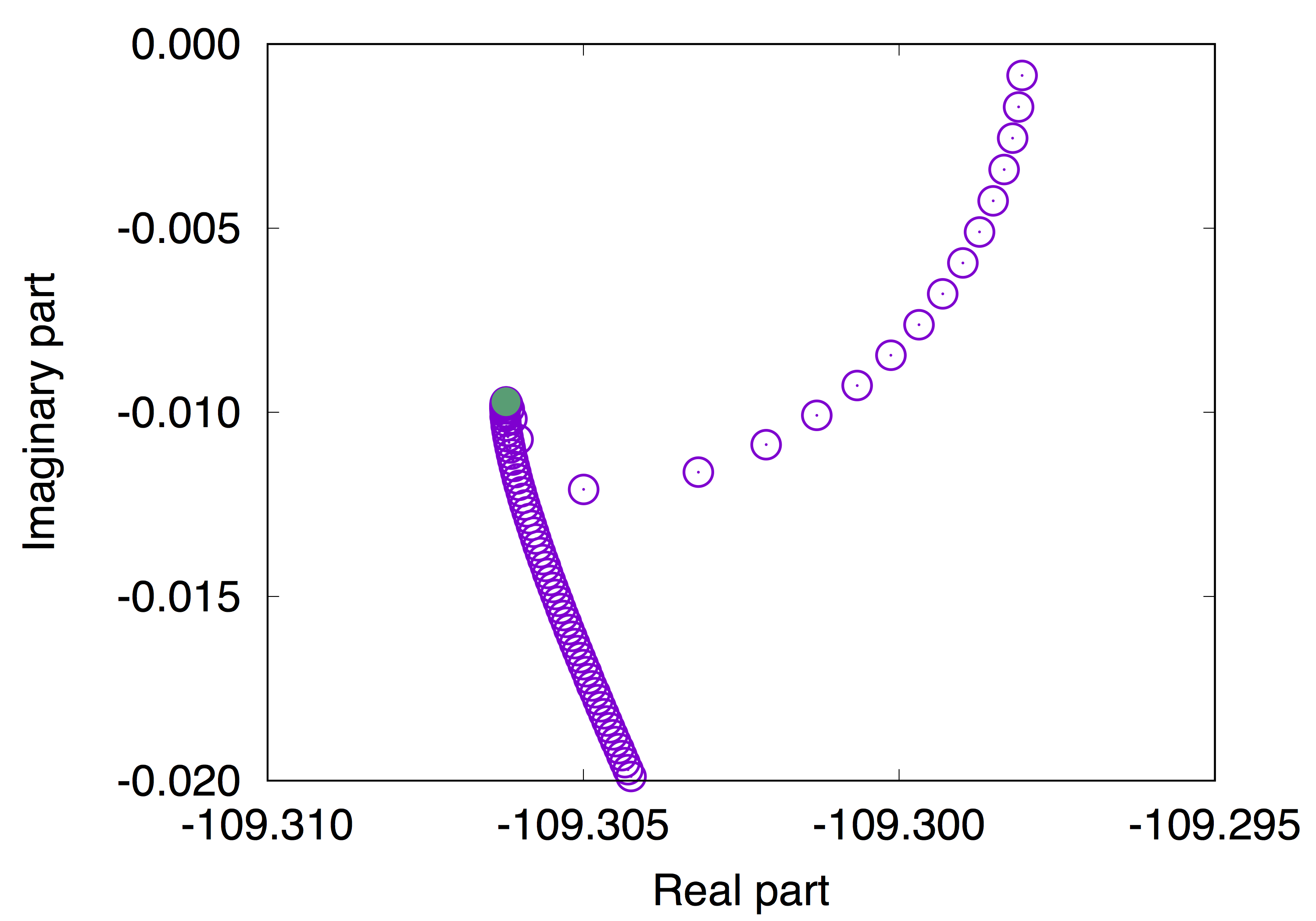}
\caption{The real and imaginary part of the energy for the resonance state. The data points correspond to $\eta = [1.0\times 10^{-5},  1.0\times 10^{-3}]$ with the interval $1.0\times 10^{-5}$. The green dot is the stationary point.\label{etafig}}
\end{figure}
We computed the XMS-CASPT2 effective Hamiltonian $H_\mathrm{eff}$ and the CAP matrix $W_\mathrm{cap}$ within the model space for N$_2$ anion ($r = 2.074$~bohr).
N$_2$ anion has been theoretically studied previously (Ref.~\onlinecite{Berman1983PRA}; see also Ref.~\onlinecite{Zuev2014JCP} and references therein).
The aug-cc-pVQZ basis set was used together with three even-tempered diffuse functions in both $s$ and $p$ shells.
The active space included 16 orbitals and 5 electrons, and
14 states were averaged in the calculation without imposing spacial symmetry.
The spherical CAP with 4.88 bohr radius was used. 
The so-called $\eta$ trajectory was obtained by diagonalizing $H_\mathrm{eff} - \eta W_\mathrm{cap}$ with various $\eta$; the result is compiled in Fig.~\ref{etafig}.
The stationary point was obtained by minimizing $\eta |dE/d\eta|$ to be $\eta = 0.000206$, $E=-109.30623-0.00971i$ (shown as a green dot in Fig.~\ref{etafig}).
Note that the electronic resonance width $\Gamma$ computed from this result (0.528~eV) is in good agreement with the experimentally derived value (0.41~eV).\cite{Berman1983PRA}

\begin{figure}[tb]
\includegraphics[width=0.48\textwidth,keepaspectratio]{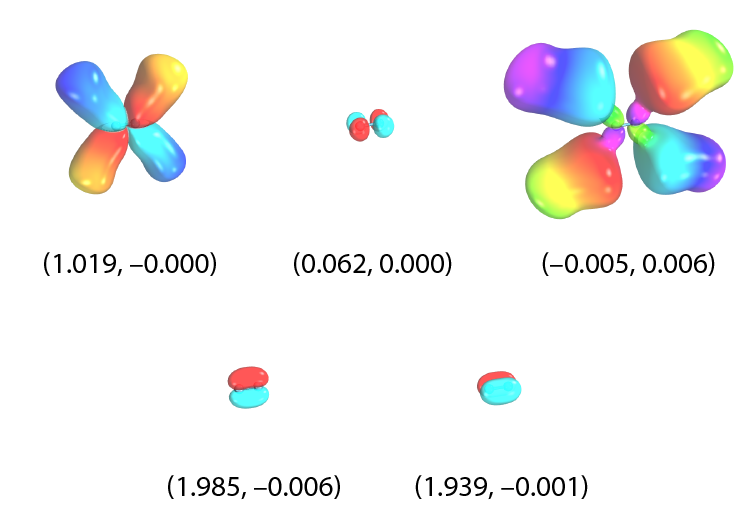}
\caption{Natural orbitals for the metastable ground state of N$_2$ anion. The numbers are (complex) occupation numbers.
The hue of the colors represents the phase of the orbitals (see Fig.~\ref{colorw}).\label{resofig}}
\end{figure}

The natural orbitals in the active space with non-zero occupation numbers are shown in Fig.~\ref{resofig}. 
We here use the convention in which the molecule is aligned with the $z$ axis, and the $x$ axis is in-plane.
There are two (almost) doubly occupied orbitals ($\pi_x$ and $\pi_y$) and a singly occupied orbital ($\pi_x^\ast$).
The $\pi_y$ and $\pi_y^\ast$ orbitals resemble very closely the neutral N$_2$ orbitals, and their occupation numbers sum up to 2. 
It is also noted that the imaginary part of the occupation numbers for $\pi_y$ and $\pi_y^\ast$ is nearly zero.
In contrast, the singly occupied orbital $\pi_x^\ast$ is partially hybridized with the diffuse orbitals that describe continuum states.
Furthermore, there is another hybridized orbital $\pi_x^{\ast\prime}$ that is a mixture of $\pi_x^\ast$ and the continuum.
The occupation number for this $\pi_x^{\ast\prime}$ orbital is $(-0.005, 0.006)$, which suggests that it plays a role in the auto-detachment process.
Quantitative interpretation of these occupation numbers requires further research.

\section{Conclusions}
In this work, we developed a software program that allows for visualizing complex-valued molecular orbitals, in which
the phases of the complex-valued orbitals are represented by the RGB color wheel.
The implementation has been on the basis of the IboView program\cite{Knizia2013JCTC,Knizia2015ACIE} and the BAGEL package.\cite{bagel}
Examples were presented for systems under a magnetic field, those with spin--orbit interactions, and those with metastable states. 
These examples have shown that the phases of the complex-valued orbitals do carry chemical and physical information (hybridization, time-reversal symmetry, and decay channels in these examples, respectively).

\begin{acknowledgments}
Professor Ksenia Bravaya is thanked for help with the calculations for metastable states.
We appreciate useful discussions with Professor Sandeep Sharma.
R.A.-S. and T.S. have been supported by Air Force Office of Scientific Research (AFOSR FA9550-18-1-0252)
and by the National Science Foundation CAREER Award (CHE-1351598), respectively.  T.S. is a Sloan Research Fellow.
\end{acknowledgments}

\end{document}